\documentclass[aps,prb,twocolumn,showpacs,preprintnumbers,final]{revtex4}

\usepackage{graphicx,amsmath,amssymb,mathrsfs,array,longtable,delarray,bm,color}

\begin{document}

\title{Kondo physics versus spin-gap physics in fully spin-split quantum wires}

\author{F. Sfigakis}
\altaffiliation{corresponding author: fs228@cam.ac.uk}
\author{C.J.B. Ford}
\author{M. Pepper}
\altaffiliation{Present address: Department of Electronic and Electrical
Engineering, University College, London, WC1E 6BT, UK.}
\author{D. A. Ritchie}
\author{I. Farrer}
\author{M. Y. Simmons}
\altaffiliation{Present address: School of Physics, University of New South Wales,
Sydney 2052, Australia.} \affiliation{Cavendish Laboratory, J. J. Thomson Avenue,
Cambridge, CB3 OHE, United Kingdom}
\author{D. Maude}
\affiliation{Grenoble High Magnetic Field Laboratory, 25 avenue des Martyrs,
Grenoble, BP 166, France}


\begin{abstract}
Linear and nonlinear transport of quantum wires are investigated at a magnetic
field where spin-split one-dimensional (1D) subbands are equidistant in energy. In
this seldom-studied regime, experiments are consistent with a density-dependent
energy gap between spin subbands, and with a complete spin polarization of the
first 1D subband under a large source-drain bias at zero field.
\end{abstract}


\maketitle

Using split-gate devices,\cite{Thornton86} the quantization of the differential
conductance $G=dI/dV_{\text{sd}}$ in units of $G_0=2e^2/h$ in ballistic quantum
wires\cite{Wharam88,vanWees88} is well understood in terms of non-interacting
electrons. However, a conductance feature near $\sim$\,0.7$\,G_0$, the so-called
0.7 anomaly or 0.7 structure,\cite{Thomas96} cannot be explained in such terms. The
discovery of ``0.7 analogs'' at Zeeman crossings,\cite{Abi03} where Zeeman-split 1D
subbands of opposite spins become degenerate in energy, initially suggested that
the physical mechanism of the 0.7 structure becomes manifest whenever two levels of
opposite spin become degenerate in energy. Despite considerable attention, both
theoretical and experimental, the physical origins of the 0.7 structure/analogs are
still unclear. Models based on spin
polarization\cite{Wang96,Thomas98-A,Nuttinck00,Kristensen00,Appleyard00-A,
Hirose01-B,Kristensen02,Reilly02,Havu04,Reilly05,Berggren05,Rokhinson06,Chiatti06,
DiCarlo06-A,Jaksch06,Lassl07,Abi07,Koop07,Berggren08,Simmonds08-A,Chen08-A,Abi08,Chen09-A}
(with an energy gap, also referred to as a ``spin gap'', opening between spin-split
subbands) and on Kondo physics\cite{Lindelof01,Cronenwett02,Meir02,Sushkov03,
Cornaglia04-A,Rejec06,Golub06,Luscher07,Aono08,Meir08,Jefferson08} (involving a
quasi-bound state in the 1D channel) can describe most -- but not all -- of the
phenomenology associated with the 0.7 structure/analogs. For example, spin
polarization models do not predict the occurrence of a zero-bias anomaly in quantum
wires,\cite{Sarkozy09-A} whilst Kondo physics models do not describe well the 0.85
plateau at high source-drain bias.\cite{Francois08-A} Also, in many experiments,
one cannot resolve which model best fits the data. For example, the shot noise
suppression near the 0.7 structure can described by either
models,\cite{DiCarlo06-A,Golub06} and the temperature dependence of the 0.7
structure can be described by either an exponential\cite{Kristensen00} or a power
law.\cite{Cronenwett02} There is thus a need for experiments that can unambiguously
point to one (or neither!) type of models.

\begin{figure}[t]
    \centering
    \includegraphics[width=1.0\columnwidth,clip]{Tdep0Tv3.eps}
    \begin{minipage}{0.30\columnwidth}
    \includegraphics[width=\columnwidth]{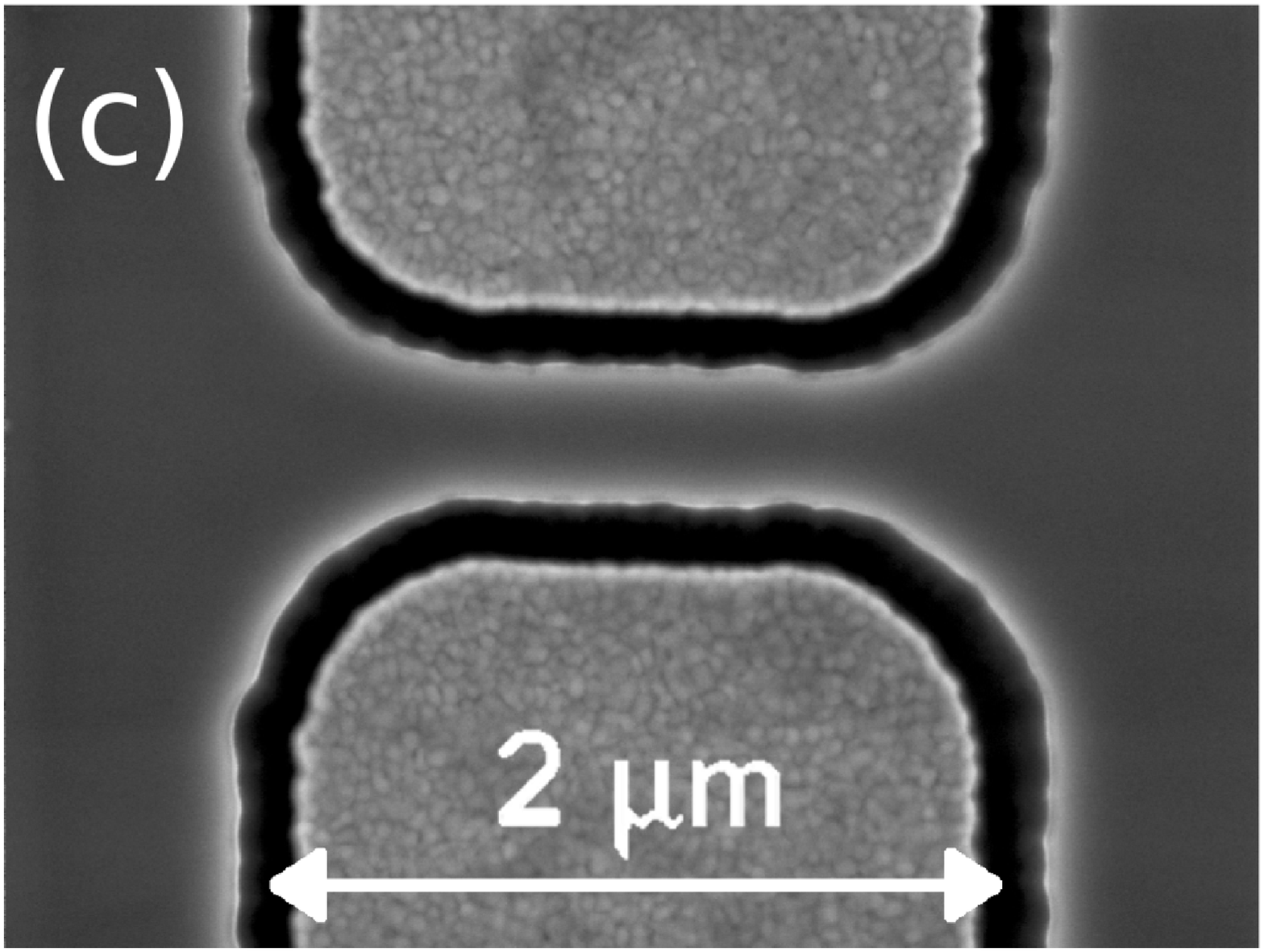}
    \end{minipage}
    \hspace{0.1cm}
    \begin{minipage}{0.50\columnwidth}
    \includegraphics[width=\columnwidth]{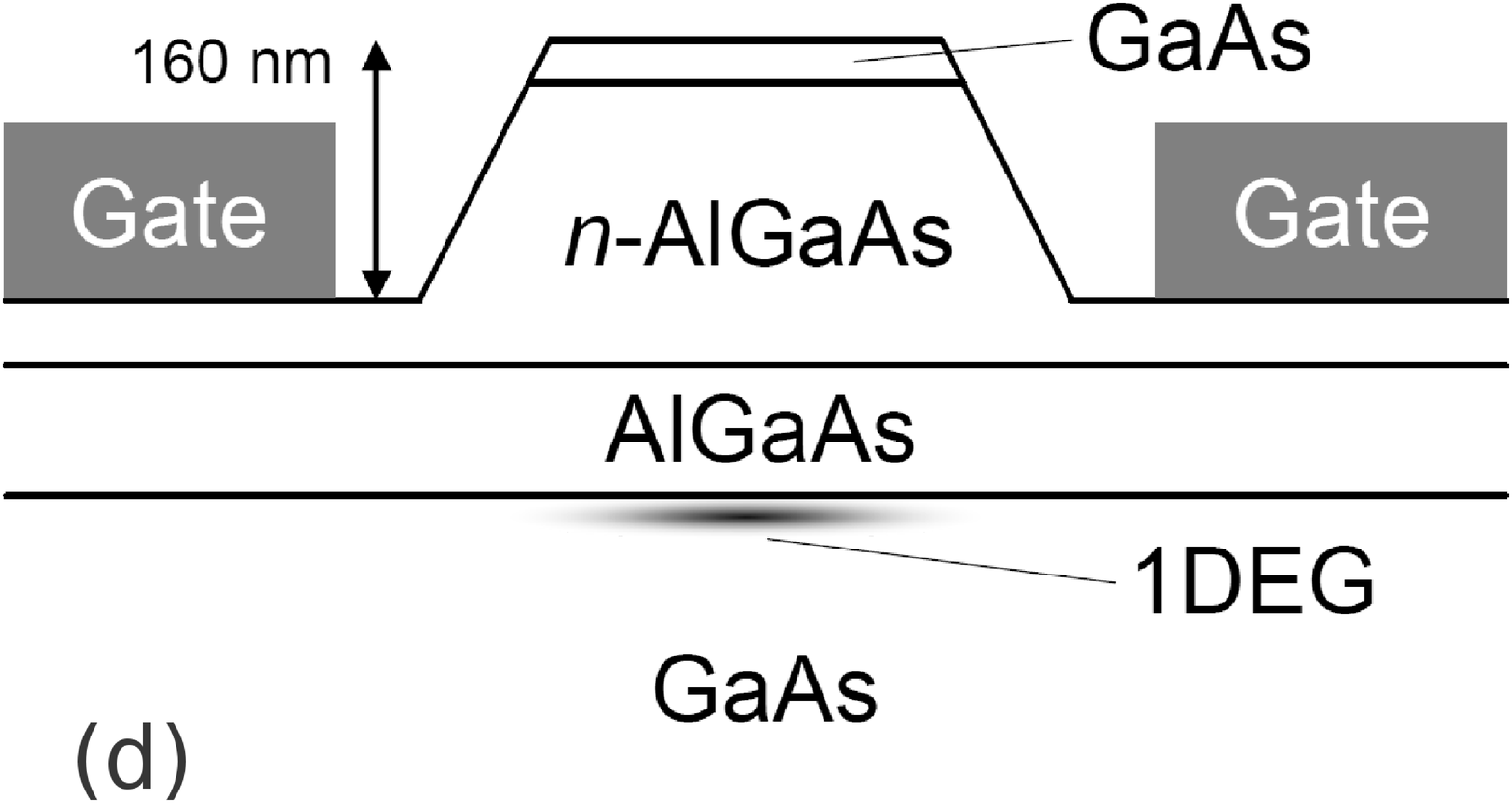}
    \end{minipage}
    \caption{(a) Measured conductance of sample A at
    $B=0$ for $T=0.3$\,K (thickest trace), 0.6\,K, and 1\,K.
    \textsc{(inset)} Energy diagram depicting the Zeeman
    effect on the first two 1D subbands. At $B_1$, the spin
    subbands are fully spin-split; at $B_2$, the $E_{\ell\uparrow}$ and
    $E_{(\ell+1)\downarrow}$ energy levels are degenerate (a ``Zeeman crossing'').
    (b) Calculated conductance using
    Eqs.~(\ref{eq:Gtheory1})--(\ref{eq:Eup0}) with
    $\omega_y/\omega_x=4$, $\hbar\omega_y=1$ meV, and
    $k_{\text{\tiny{B}}}T=0.02$ meV (thickest trace), 0.04 meV,
    0.08 meV, and 0.16 meV. In a saddle point potential, conductance quantisation
    is completely lost when $\hbar\omega_y=4k_{\text{\tiny{B}}}T$. (c) Scanning
    electron microscope (SEM) image of sample A. (d) Schematic view of the
    cross-section of the quantum wire shown in (c).}
    \label{fig:BJ56-11-Tdep0T}
\end{figure}

Transport in an in-plane magnetic field $B=B_1$ [see inset of Fig.
\ref{fig:BJ56-11-Tdep0T}(a)], where spin-split 1D subbands are equidistant in
energy ($B_1$ is halfway between $B=0$ and the Zeeman crossing at $B=B_2$), allows
the properties of spin $\downarrow$ (defined as the spin type lowest in energy at
finite $B$) and spin $\uparrow$ subbands to be studied \textit{separately}. In this
article, we demonstrate that many-body effects are not restricted merely to regimes
of near-degeneracy between electrons of opposite spin. Our experimental results are
not consistent with Kondo physics or purely pinning\cite{Nuttinck00,Kristensen02}
of the spin $\uparrow$ subband near a chemical potential. Our experimental data is
consistent to spin-gap models based on the concept of a spin gap affecting
\textit{both} spin $\uparrow$ and spin $\downarrow$ subbands.

Samples (A, B, C) were fabricated from three GaAs/AlGaAs single heterojunction 2D
electron gases (all $\sim$\,300 nm deep) with carrier densities of (1.26, 1.82,
0.94)$\times$10$^{15}$ m$^{-2}$ and mobilities of (355, 475, 195) m$^{2}$/Vs, whose
Molecular Beam Epitaxy (MBE) layer structure is (from the top): a 17\,nm GaAs cap,
a (215, 215, 200)\,nm n-doped (Si) Al$_{0.33}$Ga$_{0.67}$As doped layer, a (70, 70,
80)\,nm undoped Al$_{0.33}$Ga$_{0.67}$As spacer layer, and 500\,nm semi-insulating
GaAs. Self-aligned recessed metal gates were evaporated after a 160\,nm deep etch,
using a 1:1:38 H$_3$PO$_4$:H$_2$O$_2$:H$_2$O solution (by volume), shown in
Figs.\,1(c)-1(d). The etched 1D channels for samples (A, B, C) were (0.75, 0.02,
0.02) $\mu$m long and 0.35--0.40 $\mu$m wide. Their differential conductance was
measured in dilution refrigerators (with 0.04\,K and 0.3\,K base electron
temperatures), using standard lock-in techniques.

\begin{figure}
    \centering
    \includegraphics[width=1.0\columnwidth,clip]{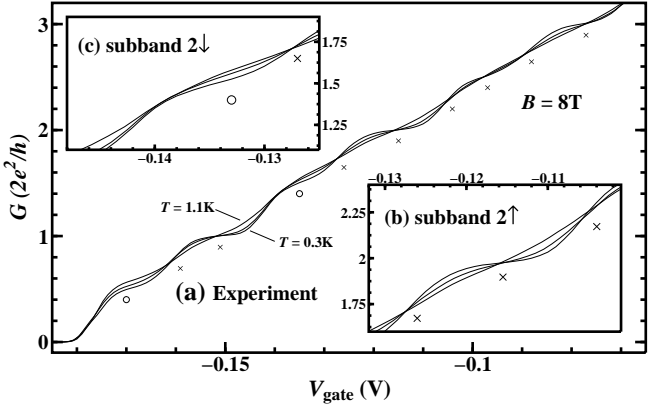}
    \caption{{(a)} Measured conductance of sample A at $B=8.0$\,T
    where spin subbands are almost fully spin-split, for $T=0.3$\,K, 0.6\,K,
    and 1.1\,K. Enlarged view of: {(b)} the integer 2.0$\,G_0$
    plateau, and {(c)} the half-integer 1.5$\,G_0$ plateau.
    The `$\times$' and `$\circ$' symbols are explained in the main text.
    \label{fig:BJ56-11-Tdep8T}}
\end{figure}

Figure \ref{fig:BJ56-11-Tdep0T}(a) shows a ``classic'' 0.7 structure in sample A as
the temperature $T$ is increased. Note the $T$-invariant points at
$G/G_0=1,~1.5,~2,~2.5,~\text{and}~3$. In an attempt to gain some physical insight,
we used a ``toy model'' to calculate conductance; it is not meant to exactly
reproduce experimental data, but rather illustrate types of behavior. Using a
saddle-point potential,\cite{Buttiker90} we calculate the 1D conductance at
equilibrium with:
\begin{eqnarray}
G(\mu,T,B)&=&\frac{e^2}{h}~\sum_{\ell=0}~
\sum_{\varsigma=\uparrow,\downarrow}~\int_{0}^{\infty}\;
\bigg[~T_{\ell\varsigma}(E)~\bigg]\qquad\nonumber\\
&&\qquad\times\left[-\frac{\partial f}{\partial E}(E,\mu,T)\right]~dE
\label{eq:Gtheory1}
\end{eqnarray}
where $T_{\ell\varsigma}(E)=[1+e^{-2\pi(E-E_{\ell\varsigma})/\hbar\omega_x}]^{-1}$
is the transmission coefficient for each spin subband, $E_{\ell\varsigma}$ are
given by
\begin{eqnarray}
E_{\ell\downarrow}&=&\hbar\omega_{\!y}(\ell+\frac{_1}{^2})\,-\,
\frac{_1}{^2}|g|\mu_{\text{\tiny{B}}}B \label{eq:Edn0}\\
E_{\ell\uparrow}&=&\hbar\omega_y(\ell+\frac{_1}{^2})\,+\,
\frac{_1}{^2}|g|\mu_{\text{\tiny{B}}}B, \label{eq:Eup0}
\end{eqnarray}
$\varsigma$ labels the spin type ($\uparrow,\downarrow$), $\ell$ the 1D subband
index, $\mu$ is the equilibrium chemical potential, $k_{\text{\tiny{B}}}$ the
Boltzmann constant, $-\frac{\partial f}{\partial E}(E,\mu,T)=[4k_{\text{\tiny{B}}}
T\cosh^2[(E-\mu)/2k_{\text{\tiny{B}}}T]]^{-1}$ the derivative of the Fermi
function, $\hbar\omega_x$ the energy broadening due to quantum tunneling,
$\hbar\omega_y$ the 1D subband energy level spacing, $g$ the bulk GaAs Land\'{e}
$g$-factor ($|g|=0.44$), and $\mu_{\text{\tiny{B}}}$ the Bohr magneton. In both
calculations and experiments, magnetic field $B$ is in-plane, along the current
flow through the 1D channel in the $x$-direction.  The $y$-direction is in-plane,
perpendicular to the current flow, and the $z$-direction is out of the plane. For
simplicity, the effects of diamagnetic shift\cite{Stern68,Schuh85,Salis99-B} have
been omitted. Provided the pinch-off voltage does not drift with
time\cite{TdepTimedrift} and the 1D constriction can be described by a saddle-point
potential, $T$-invariant points should occur either (i) at mid-height of a riser
whenever a subband energy level $E_{\ell\varsigma}$ crosses $\mu$, or (ii) at
mid-length of a plateau whenever $\mu$ is exactly halfway between two subband
energy levels. Comparing Figs.\,1(a) with 1(b), the conductance $G$ behaves mostly
as predicted by Eq.~(\ref{eq:Gtheory1}) for $G \geqslant 2e^2/h$. The same
observations apply to split-gated samples, e.g. see Figure 1(a) in
Ref.\,\onlinecite{Cronenwett02} and Figure 4 in Ref.\,\onlinecite{Thomas96}.

\begin{figure}[t]
    \centering
    \includegraphics[width=1.0\columnwidth,clip]{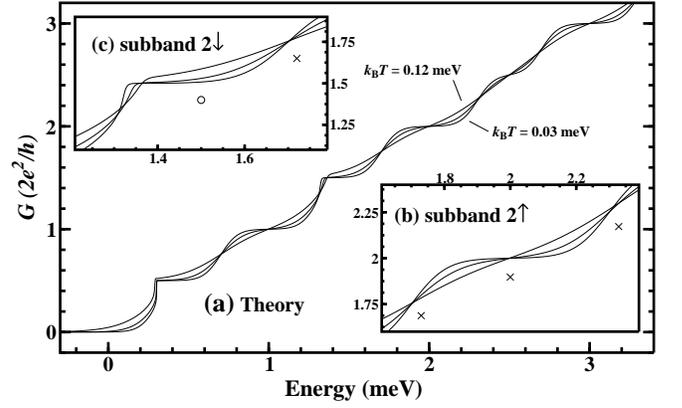}
    \caption{{(a)} Simulated conductance at $|g|\mu_{\text{\tiny{B}}}B/\hbar\omega_y=0.4$,
    using Eqs.(\ref{eq:Gtheory1})--(\ref{eq:1Ddensity}) with
    $\gamma_0=60$, $\gamma_1=30$, and $\gamma_2=0$\,$\mu$eV$\cdot\mu$m$^2$,
    $\omega_y/\omega_x=4$, $\hbar\omega_y=1$\,meV, $r=2$, and
    $k_{\text{\tiny{B}}}T=0.03$, 0.06, and 0.12\,meV.
    Enlarged view of: {(b)} the integer 2.0$\,G_0$
    plateau, and {(c)} the half-integer 1.5$\,G_0$ plateau.
    \label{fig:G1DtryH04}}
\end{figure}

Figure \ref{fig:BJ56-11-Tdep8T} shows the (constant-voltage) four-terminal
conductance of sample A for the first six almost fully spin-split plateaus ($B
\lesssim B_1$). The riser of each quantized plateau is due to the population of the
1$\downarrow$, 1$\uparrow$, 2$\downarrow$, 2$\uparrow$, 3$\downarrow$, and
3$\uparrow$ spin subbands, in that order. The measured conductance has many
$T$-invariant points (indicated by a `$\times$'), as predicted by
Eq.~(\ref{eq:Gtheory1}). However, there are none at mid-length of the 0.5$\,G_0$
and 1.5$\,G_0$ quantized plateaus: as $T$ increases, almost all of the plateau
rises in conductance (indicated by an `$\circ$'). This has been previously observed
in split-gated samples,\cite{Thomas02,Abi03} but only for the spin-split 0.5$\,G_0$
plateau and no mechanism was proposed. If $|g|\mu_{\text{\tiny{B}}}B \!<\!
4k_{\text{\tiny{B}}}T \!<\! 2\hbar\omega_y$ in the regime $0\!<\!B\!<\!B_1$,
Eq.~(\ref{eq:Gtheory1}) predicts the $T$-invariant points at 0.5, 1.0, 1.5, 2.0,
2.5, and 3.0$\,G_0$ always remain visible whilst those at 0.75, 1.75, and
2.75$\,G_0$ only disappear if $\hbar\omega_y \!<\! 4k_{\text{\tiny{B}}}T$. Clearly,
the latter $T$-invariant points are still present in Figure
\ref{fig:BJ56-11-Tdep8T} at all temperatures: the observed thermal broadening
involves at most only one spin subband at mid-height of each riser, or at most only
two spin subbands on the middle of each plateau. Furthermore, the 2.5$\,G_0$
plateau behaves as expected from non-interacting electrons, with a $T$-invariant
point at 2.5$\,G_0$. The rise of the 1.5$\,G_0$ and 0.5$\,G_0$ plateaus above their
nominal quantized value with increasing temperature cannot be explained within a
single-particle picture.

At $B=0$, the rise of the 0.7 structure to $G_0=2e^2/h$ with decreasing temperature
has been suggested to result from Kondo physics.\cite{Cronenwett02} At $B=8$ T, the
behavior of the 0.5$\,G_0$ and 1.5$\,G_0$ plateaus in Fig.~\ref{fig:BJ56-11-Tdep8T}
cannot be attributed either to Kondo physics: the experimental $G$ rises with
increasing $T$ (the opposite temperature dependence is
predicted\cite{Meir02,Matveev04-B}). Furthermore, the Kondo effect is completely
suppressed in large magnetic fields (far away from spin degeneracy points at $B=0$
and $B=B_2$), such as when the spin-split 0.5$\,G_0$ plateau appears. Pinning alone
of the spin $\uparrow$ subbands to the chemical potential\cite{Nuttinck00} also
cannot explain the behaviour of the 0.5$\,G_0$ and 1.5$\,G_0$ plateaus associated
with populating spin $\downarrow$ subbands. However, the behavior shown in Figure 2
could be consistent with a density-dependent spin gap opening between spin
subbands.

To illustrate this, we calculated the conductance using Eq.~(\ref{eq:Gtheory1}),
but with $E_{\ell\downarrow}$ redefined as (valid only if $B \approx B_1$, see
further below):
\begin{equation}
E_{\ell\downarrow}\approx\hbar\omega_y(\ell+\frac{_1}{^2})
-\gamma_\ell(n_{\ell\downarrow}+n_{\ell\uparrow})^r
\label{eq:Edown}
\end{equation}
where $(n_{\ell\downarrow}+n_{\ell\uparrow})$ is the electron density of 1D subband
$\ell$ (per unit length along the 1D channel) and $\gamma_\ell$ is the
electron-electron interaction strength in subband $\ell$. The individual subband
densities $n_{\ell \varsigma}$ are calculated using:
\begin{equation}
n_{\ell\varsigma}~=~
\int_{E_{\ell\varsigma}}^{\infty}g_{\textsc{1d}}(E,E_{\ell\varsigma}) ~f(E,\mu,T)~dE
\label{eq:1Ddensity}
\end{equation}
where
$g_{\textsc{1d}}(E,E_{\ell\varsigma})=\sqrt{2m^*}(2\pi\hbar\sqrt{E-E_{\ell\varsigma}})^{-1}$
is the 1D density of states. All variables $E_{\ell\varsigma}$ and
$n_{\ell\varsigma}$ were calculated self-consistently. It is unlikely that a spin
gap occurring in real samples increases indefinitely with increasing electron
density: it must eventually either saturate or close.\cite{Wang96,Jaksch06}
However, when $E_{\ell\downarrow}$ is deep below $\mu$, the $\ell\!\!\downarrow$
spin subband contributes $e^2/h$ to the total conductance, regardless of whether
the spin gap is open or closed. Therefore, for simplicity, we did not include any
term in Eq.~(4) to ensure that the spin gap eventually closes or saturates. At
$B=0$ or near Zeeman crossings ($B\approx B_2$), in many spin-gap models, the
apparent ``pinning'' of $E_{\ell\uparrow}$ near a chemical potential is thought
responsible for the appearance of the 0.7 structure/analogs. However, our data
(Figs.~2 and 4) do not show unusual behavior associated with $E_{\ell\uparrow}$
when $B\approx B_1$: either $E_{\ell\uparrow}$ is far enough above $\mu$ to
contribute very little to the total conductance regardless of any spin gap, or the
spin gap must have already closed or saturated when $E_{\ell\uparrow}$ crosses
$\mu$. Again for simplicity, we used Eq.~(3) to describe $E_{\ell\uparrow}$. These
simplifying approximations are only valid if $B\approx B_1$.

The magnetic field in Figure \ref{fig:G1DtryH04} is
$|g|\mu_{\text{\tiny{B}}}B/\hbar\omega_y=0.4$ (note $B\!\equiv \!B_1$ if
$|g|\mu_{\text{\tiny{B}}}B/\hbar\omega_y=\frac{1}{2}$) to be closer to the
experimental situation in Figure 2. The simulated 0.5$\,G_0$ and 1.5$\,G_0$
half-integer plateaus rise above their nominal value, while all other plateaus
behave as expected from a single-particle picture. The exact functional form of the
opening spin gap is not critical: similar behavior was also obtained for
simulations with $r=\frac{_1}{^2}$ and $r=1$. The key here is that spin
$\downarrow$ subbands populate much more rapidly below the chemical potential than
spin $\uparrow$ subbands.

\begin{figure}
    \centering
    \includegraphics[width=1.0\columnwidth,clip]{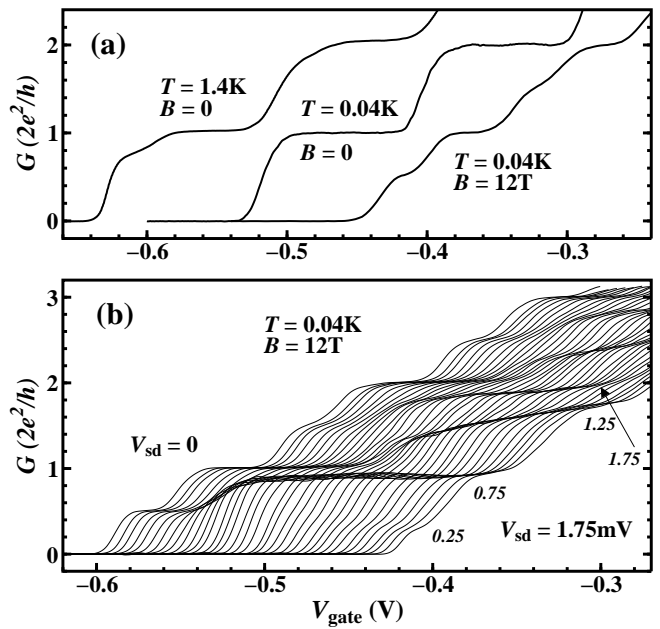}
    \caption{{(a)} Measured conductance at equilibrium of sample
    B in three different regimes (traces offset laterally).
    {(b)} Non-equilibrium conductance in magnetic field
    from $V_{\!\text{sd}}=0$ to $V_{\!\text{sd}}=+1.75$ mV in 0.05 mV steps
    (traces offset laterally). Note the evolution from half-integer plateaus
    to quarter-integer plateaus as $V_{\!\text{sd}}$ increases.
    \label{fig:Vdep12T}}
\end{figure}

Figure \ref{fig:Vdep12T}(a) shows a ``classic'' 0.7 structure at 1.4 K in sample B,
with a clean $2e^2/h$ plateau at 0.04 K. Although $B_1>12$~T, the 1D subbands are
clearly spin-split. Figure \ref{fig:Vdep12T}(b) shows conductance traces under
increasing $V_{\!\text{sd}}$ at $B=12$~T, where so-called ``quarter-integer''
plateaus\cite{Patel91-B,Thomas98-B} are observed. For subband $\ell=2$, the
2.75$\,G_0$ and 2.25$\,G_0$ quarter-integer plateaus are easily identifiable, with
the latter appearing at a slightly larger $V_{\!\text{sd}}$ than the 2.75$\,G_0$
plateau. For subband $\ell=1$, the 1.75$\,G_0$ plateau has almost fully formed
while the 1.25$\,G_0$ plateau has only begun to form. For subband $\ell=0$, the
trend continues with a very well defined plateau at $\sim$ 0.8$\,G_0$ but with no
signs of a 0.25$\,G_0$ plateau until a very large $V_{\!\text{sd}}$ is applied.
Identical behavior occurs for $V_{\!\text{sd}}<0$ (not shown). In high-indexed
plateaus ($\ell\!\geqslant\!4$) where one expects electron-electron interactions to
be minimal, both the ($\ell+0.25$)$\,G_0$ and ($\ell+0.75$)$\,G_0$ plateaus appear
at the \textit{same} $V_{\!\text{sd}}$ (not shown). Kondo physics cannot account
for this behavior. However, the behavior shown in Figure 4 could be consistent with
a density-dependent spin gap opening between spin subbands.

\begin{figure}
    \centering
    \includegraphics[width=1.0\columnwidth,clip]{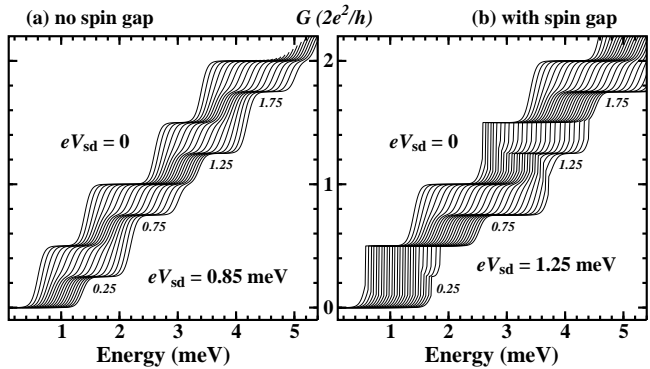}
    \caption{Simulated conductance traces using Eq.(\ref{eq:Gtheory2})
    for a fixed $V_{\!sd}$ at $k_{\text{\tiny{B}}}T=5$ $\mu$eV and
    $|g|\mu_{\text{\tiny{B}}}B=0.4\hbar\omega_y$ for a 1D channel
    with $\omega_y/\omega_x=4$ and $\hbar\omega_y=2$ meV
    in two situations: {(a)} non-interacting electrons, and
    {(b)} interacting electrons. Source-drain bias $V_{\!\text{sd}}$ is increased in
    0.05 meV steps from left to right (traces offset laterally). Note the late onset
    of the 0.25$\,G_0$ and 1.25$\,G_0$ quarter-integer plateaus in
    (b) compared to (a). Parameters used in the calculations
    were $r=1$, $\gamma_0=150$ $\mu$eV$\cdot\mu$m, $\gamma_1=75$ $\mu$eV$\cdot\mu$m,
    and $\gamma_2=0$. Variables $E_{\ell\varsigma}$ and
    $n_{\ell\varsigma}$ were calculated self-consistently using Eqs.
    (\ref{eq:Edown})--(\ref{eq:1Ddensity}).
    \label{fig:VsdBcalc}}
\end{figure}

\begin{figure}[h]
    \centering
    \includegraphics[width=1.0\columnwidth,clip]{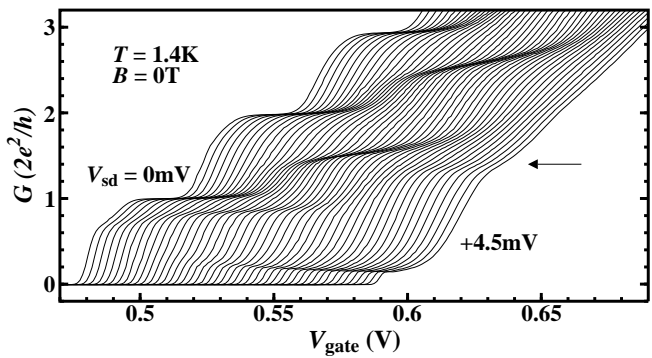}
    \caption{Measured conductance of sample C
    at $B=0$T (traces offset laterally) from $V_{\!\text{sd}}=0$ (leftmost) to
    $V_{\!\text{sd}}=+4.5$ mV (rightmost) in 0.1 mV steps. Note how the
    $\sim\,$0.8$\,G_0$ and the $\sim\,$0.3$\,G_0$ plateaus look nearly identical
    to the 0.75$\,G_0$ and 0.25$\,G_0$ plateaus of Figure \ref{fig:Vdep12T}(b),
    except for the presence of the 0.5$\,G_0$ plateau near $V_{\!\text{sd}}=0$.
    \label{fig:Vdep0T}}
\end{figure}

Applying the Glazman-Khaetskii formalism\cite{Glazman89-B} to
Eq.~(\ref{eq:Gtheory1}), the non-equilibrium 1D conductance becomes:
\begin{eqnarray}
G(\mu,T,B,V_{\!\text{sd}})&=&\frac{e^2}{h}~\sum_{\ell=0}~
\sum_{\varsigma=\uparrow,\downarrow}~\int_{0}^{\infty}
\bigg[-\frac{\partial f}{\partial E}(E,\mu,T)\bigg]\nonumber\\
&&\times\,\frac{1}{2}\bigg[~T_{\!\ell\varsigma}(\mu_s)\,+\,
T_{\!\ell\varsigma}(\mu_d)\,\bigg]~dE~~ \label{eq:Gtheory2}
\end{eqnarray}
where $\mu_s \!=\! E+\frac{1}{2}eV_{\!\text{sd}}$ and $\mu_d \!=\!
E-\frac{1}{2}eV_{\!\text{sd}}$ are the chemical potentials at source and drain
respectively. Figure \ref{fig:VsdBcalc}(a) shows calculated non-equilibrium
conductance traces for non-interacting electrons at a fixed magnetic field. Since
$|g|\mu_{\text{\tiny{B}}}B/\hbar\omega_y<\frac{1}{2}$, the spin-split half-integer
plateaus disappear before the integer plateaus (as in the experimental data). Note
how the onsets of the ($\ell+0.25$)$\,G_0$ and ($\ell+0.75$)$\,G_0$ quarter-integer
plateaus appear at the \textit{same} $V_{\!\text{sd}}$. By contrast, with a spin
gap, Figure \ref{fig:VsdBcalc}(b) shows that the 0.25$\,G_0$ and 1.25$\,G_0$
plateaus appear at a much \textit{higher} $V_{\!\text{sd}}$ than their 0.75$\,G_0$
and 1.75$\,G_0$ counterparts (as in the experimental data). Essentially, the
($\ell+0.25$)$\,G_0$ plateau can only form once $eV_{\!\text{sd}}$ exceeds the
spin-gap energy $\gamma_\ell(n_{\ell\downarrow}+n_{\ell\uparrow})^r$ of the
$\ell\!\downarrow$ spin subband.

It is interesting to compare Figures 4 and 5 with Figure \ref{fig:Vdep0T}, showing
the effect of a finite $V_{\!\text{sd}}$ on sample C at $B=0$. For $G\!>\!2e^2/h$,
plateaus at half-integer multiples of $\,G_0$ appear,\cite{Patel91-A} as expected
in a single-particle picture. For $G\!<\!2e^2/h$, plateaus are observed at
0.8--0.9$\,G_0$ and 0.2--0.3$\,G_0$, which have often been assigned the values
0.85$\,G_0$ and 0.50$\,G_0$ in the
literature.\cite{Kristensen00,Cronenwett02,Reilly06} These have been shown not to
be associated with Kondo physics in samples where a bound state was deliberately
enginnered to form in the 1D channel.\cite{Francois08-A} Here, we essentially
reproduce the same result in clean quantum wires (of which sample C is
representative). Furthermore, the zero-field 0.8--0.9$\,G_0$ and 0.2--0.3$\,G_0$
plateaus do not show any signs of splitting as $B$ is
increased,\cite{Thomas98-B,Abi06} leading us to interpret them as zero-field
quarter-integer plateaus, nominally at 0.75$\,G_0$ and
0.25$\,G_0$.\cite{Simmonds08-A,Chen08-A}  This would be consistent with the
smulations of Fig.\,5(b), where the spin-gap energy,
$\gamma_\ell(n_{\ell\downarrow}+n_{\ell\uparrow})^r$, increases with increasing
source-drain bias $V_{\!\text{sd}}$.

Consequently to the reasoning above, the plateau near $\sim$1.4$\,G_0$ at the
highest $V_{\!\text{sd}}$ in Fig.~\ref{fig:Vdep0T} (see arrow), also reported in
Refs.~\onlinecite{Kristensen00} and \onlinecite{Reilly05}, should be the
1.25$\,G_0$ quarter-integer plateau. Unambiguous identification of plateaus at
finite $V_{\!\text{sd}}$ has always been difficult because of the rise in $G$ with
increasing $V_{\!\text{sd}}$. This rise can be described by
``self-gating'',\cite{Kristensen00} a single electron effect.

In conclusion, we have presented data showing interacting electron effects in
finite magnetic fields, far away from near-degeneracy points between spin-split
subbands, consistent with the existence of a spin gap and inconsistent with Kondo
physics. We have also demonstrated, in a clean quantum wire, that the so-called
0.85$\,G_0$ plateau does not result from the suppression of a Kondo-enhanced
conductance from the $2e^2/h$ plateau, but is rather a fully spin-split plateau.
Within a spin-gap interpretation of the 0.7 structure, our results strongly suggest
that spin $\downarrow$ subbands are also affected by the spin gap, a mechanism that
had not been made experimentally evident until now. Finally, we propose to use
quantum wires under a large source-drain bias (in a regime where the 0.25$\,G_0$
plateau becomes visible) to produce a uni-directional, spin-polarised current.

We thank A.C. Graham, V. Tripathi, T.-M. Chen, K.-F. Berggren, and K. J. Thomas for
insightful discussions, and K. Cooper, A. Beckett, R. Milanole and L. Lozano for
technical assistance. This work was supported by the EPSRC (UK).

\bibliographystyle{apsrev}

\begin{thebibliography}{58}
\bibitem[{Tde()}]{TdepTimedrift} \bibinfo{note}{During a temperature dependence,
    one can easily diagnose
  drifting in time by comparing traces obtained whilst increasing and
  decreasing the temperature.}
\end{thebibliography}

\end{document}